\documentclass{article}
\usepackage{spconf,amsmath,graphicx}
\usepackage{hyperref}

\title{Respiratory Distress Detection from Telephone Speech using Acoustic and Prosodic Features}
\name{Meemnur Rashid$^1$, Kaisar Ahmed Alman$^1$, Khaled Hasan$^2$, John H.L. Hansen$^3$, Taufiq Hasan$^1$}
\address{
  $^1$mHealth Lab, Department of Biomedical Engineering, BUET, Dhaka - 1205, Bangladesh.\\ 
  $^2$Digital Healthcare Solutions, Dhaka - 1229, Bangladesh.\\
  $^3$Center for Robust Speech Systems (CRSS), The University of Texas at Dallas, Richardson, TX 75080.\\ 
  }

\begin{document}
\maketitle
\begin{abstract}
With the widespread use of telemedicine services, automatic assessment of health conditions via telephone speech can significantly impact public health. This work summarizes our preliminary findings on automatic detection of respiratory distress using well-known acoustic and prosodic features. Speech samples are collected from de-identified telemedicine phonecalls from a healthcare provider in Bangladesh. The recordings include conversational speech samples of patients talking to doctors showing mild or severe respiratory distress or asthma symptoms. We hypothesize that respiratory distress may alter speech features such as voice quality, speaking pattern, loudness, and speech-pause duration. To capture these variations, we utilize a set of well-known acoustic and prosodic features with a Support Vector Machine (SVM) classifier for detecting the presence of respiratory distress. Experimental evaluations are performed using a 3-fold cross-validation scheme, ensuring patient-independent data splits. We obtained an overall accuracy of 86.4\% in detecting respiratory distress from the speech recordings using the acoustic feature set. Correlation analysis reveals that the top-performing features include loudness, voice rate, voice duration, and pause duration.  
\end{abstract}
\section{Introduction}
\label{sec:intro}

Respiratory diseases, including asthma, chronic obstructive pulmonary disease (COPD), acute lower respiratory tract infection, and lung cancer \cite{marciniuk2017global}, are among the leading causes of death and disability worldwide.  According to WHO, about 339 million people suffer from asthma \cite{asthma}. COPD was responsible for about 3.17 million deaths in 2015, and in 2016, about 251 million people were affected by this disease \cite{copd}. 
Cystic fibrosis, another respiratory disease, also called Bronchiectasis, affects about 30,000 people in the US \cite{Top8}. Shortness of breath, stubborn cough, wheezing, chest pain are typical symptoms of respiratory diseases \cite{leach2008symptoms}. The COVID-19 pandemic, first identified in late 2019, is also a respiratory disease \cite{chiu2005human} that has already claimed over 1 million lives worldwide. 
Early detection of symptoms such as respiratory distress can be vital in tracking various respiratory diseases, including the COVID-19 pandemic.

Early diagnosis is critical in effectively treating most respiratory diseases, including COPD and asthma \cite{van2003detection}. Spirometry is the most commonly used test for an initial respiratory assessment of a patient. Confirmatory diagnosis of various respiratory disorders may require additional examination such as a chest x-ray or other laboratory tests. 
It is well known that shortness of breathing affects the human speech production mechanism \cite{tehrany2015speech} and, in theory, respiratory distress should be detectable from the speech signal analysis alone. Thus, automatic initial assessment of respiratory function from speech can be valuable in low-resource settings where there is a lack of experienced general practitioners.

Previous work on speech signal analysis for respiratory assessment has been minimal. Relevant research in this area includes the detection of breathing sound \cite{ruinskiy2007effective}, wheezing \cite{li2017design} and cough \cite{tracey2011cough}. In \cite{nallanthighal2019deep}, methods are presented for estimating and visualizing the breathing pattern from speech recordings. Although the authors claim that the system can be used for detecting pathological conditions, only healthy volunteer data were used for evaluation. In \cite{goel2016spirocall}, the authors introduced ``Spirocall", a method of performing spirometry using standard telephone calls. This method uses a 3D printed whistle and transmits the breathing effort's sound via the telephone channel. Although the method is promising, it still requires a 3D printed device and the presence of trained personnel to be effective. 

In this work, we present our preliminary findings on detecting respiratory distress (or shortness of breath) from conversational telephone speech. Voice recordings were collected from a telemedicine provider while the patient's personal identifying information was removed. We hypothesized that respiratory distress would affect the speech sound and rhythm and therefore utilized a set of acoustic and prosodic features for classification between patients having respiratory distress and healthy subjects. 

\section{Background}
Speech production involves airflow from the lungs through the larynx, vocal cord vibration, and resonance in the oral and nasal cavities. Since the lung is a vital organ for speech production, and respiratory disease is expected to cause physical changes in the speech production pathway and affect speech signal itself \cite{mohamed2014voice}. 
\begin{figure}[t]
  \centering
  \includegraphics[trim=0 0 0 0,clip,width=0.8\linewidth]{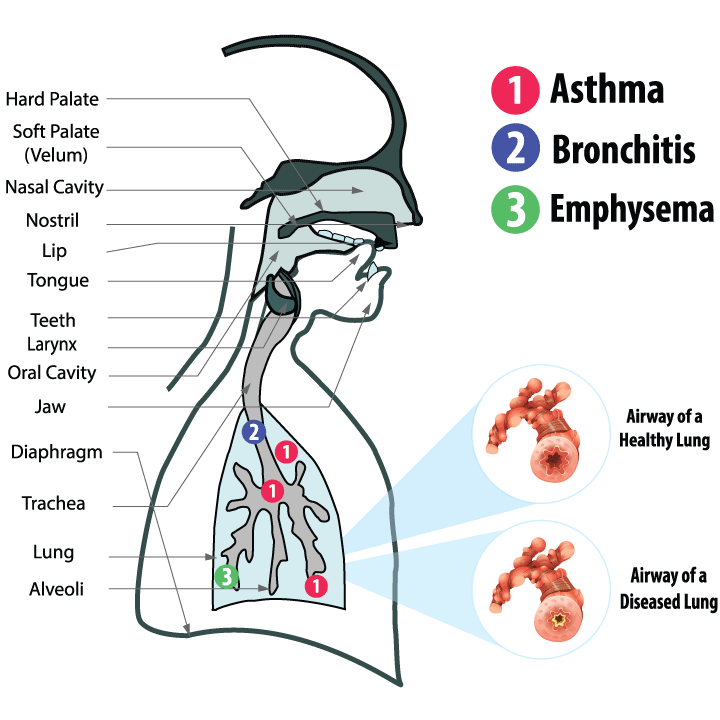}
  \caption{Effect of different respiratory diseases affect the speech production system.} 
  \label{fig:7}
\end{figure}
Previous research \cite{lee1993speech} shows that the lung volumes and breathing patterns during speech differ from quiet respiration, and alterations in speech breathing are disease and task-specific. Asthmatic patients tend to show an increased duration of silence between speech segments, lower syllable rates, and increased time in non-speech ventilatory activity \cite{doi:10.1044/jshd.5302.186}. Thus, we can assume that respiratory distress will affect the speaking rate and speech breathing pattern \cite{tehrany2015speech}.
In \cite{kutor2019speech}, a high degree of correlation was found between FEV1/FEV ratio obtained from spirometry and harmonics-to-noise ratio (HNR) obtained from human speech for asthma patients. This motivates us to consider traditional acoustic features to analyze speech signals to detect respiratory distress symptoms.

\section{Dataset}
\label{sec:dataset}
\subsection{Telemedicine audio recordings}
The speech recordings used for this study are collected from Digital Healthcare Solutions (DHS), a leading telemedicine service provider in Bangladesh\footnote{The study was approved by the clinical advisory board of DHS on February 19, 2019 (Ref \# CAB/DHS/GTT/1/2019/2/19).}. 
The telemedicine service operates through direct phone calls between the patient and physicians. For this study, a total of 88 phone-call recordings are collected. 
A patient's recordings were included if the patient called in to report suffering from respiratory distress. The speech recordings can be categorized into three classes: (i) patients reporting severe respiratory distress who are advised to visit a hospital urgently, (ii) patients reporting mild respiratory distress and generally have a history of breathing difficulty, (iii) healthy control subjects.
The data is summarized in Table~\ref{tab:2}. 
Patients included in (i) and (ii) may have a chronic condition such as asthma. However, our study's goal is to detect the condition of respiratory distress (or shortness of breath), not the actual disease. For this reason, the disease information is not used for our analysis even if it is available for some patients.
The recordings are collected at a sampling rate of 44.1 kHz.

\subsection{Annotation of the recordings}
\begin{figure}[t]
  \centering
  \includegraphics[width=\linewidth]{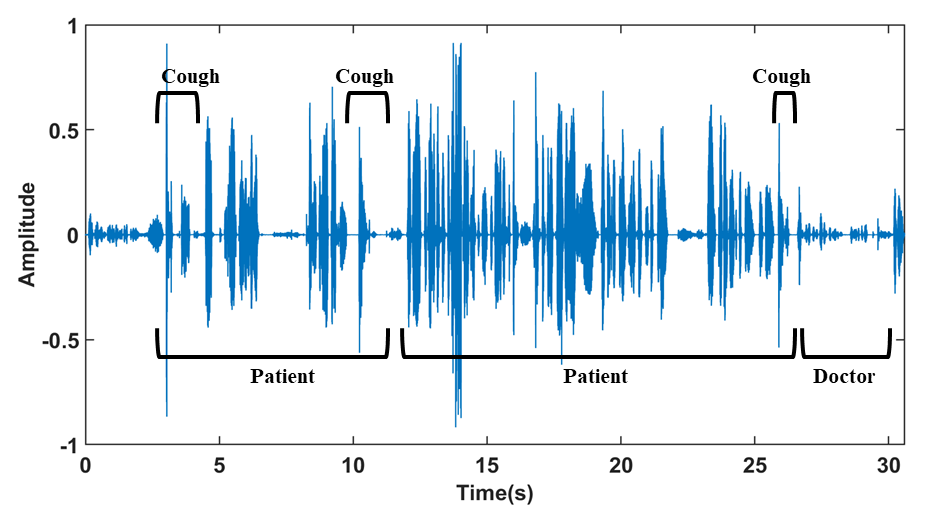}
  \caption{A typical telephone call from the dataset containing speech and cough sounds.}
  \label{fig:1}
\end{figure}

Every telemedicine phonecall consists of the speech of the patient (or their representative, e.g., relative or guardian) and the physician. The recordings also contain occasional cough and wheezing sounds from the patient or other background noise.
Accordingly, the audio recordings are annotated in these 6 categories: (i) patient, (ii) doctor, (iii) patient representative, (iv) cough sound, (v) wheezing sound, and (vi) background noise. A typical annotated phonecall recording is shown in Fig. \ref{fig:1} including some of these audio events. 
The total number of annotated recordings from all categories is 4522 as summarized in Table \ref{tab:1}. In this work, only male patients are considered. The age of the patients range between 25 - 65. A total of 1957 utterances (Table~\ref{tab:2}) of different lengths were prepared for cross-validation experiments.
\begin{table}[ht]
  \caption{Number of patients and healthy subjects and corresponding number of annotated speech segments.}
  \label{tab:2}
  \centering
  \small
  \begin{tabular}{lcc}
    \hline
    \multicolumn{1}{c}{\textbf{Condition of subjects}} &  \multicolumn{1}{c}{\textbf{No.}} &  \multicolumn{1}{c}{\textbf{Segments}}\\
    \hline
    Patient in severe respiratory distress    &5 &65~~~               \\
    Asthma patient in mild respiratory distress &5 &216 ~~~              \\
    Patient in mild respiratory distress      &26 &673~~~              \\
    Healthy subjects with no respiratory distress &52 &1003~~~              \\
    \hline
    Total &88 & 1957~~~\\
    \hline
  \end{tabular}
\end{table}

\begin{table}[ht]
  \caption{Summary of annotated speech/audio segments}
  \label{tab:1}
  \centering
  \begin{tabular}{lcc}
    \hline
    \multicolumn{1}{c}{\textbf{Category}} &  \multicolumn{1}{c}{\textbf{Segments}} & \multicolumn{1}{c}{\textbf{Duration}}\\
    \hline
    Patient                         &1957&2:10:46~~~             \\
    Doctor                          &1839&2:47:10~~~             \\
    Spokesperson                    &150&0:09:50~~~              \\
    Cough                           &88&0:01:04~~~               \\
    Wheezing                        &114&0:01:35~~~              \\
    Background Noise                &374&0:06:56~~~              \\
    \hline
    Total & 4522&5:17:21~~~\\
    \hline
  \end{tabular}
\end{table}

\subsection{Data preparation for cross-validation}
\begin{figure}[b]
  \centering
  \includegraphics[trim=50 145 50 150,clip,width=\linewidth]{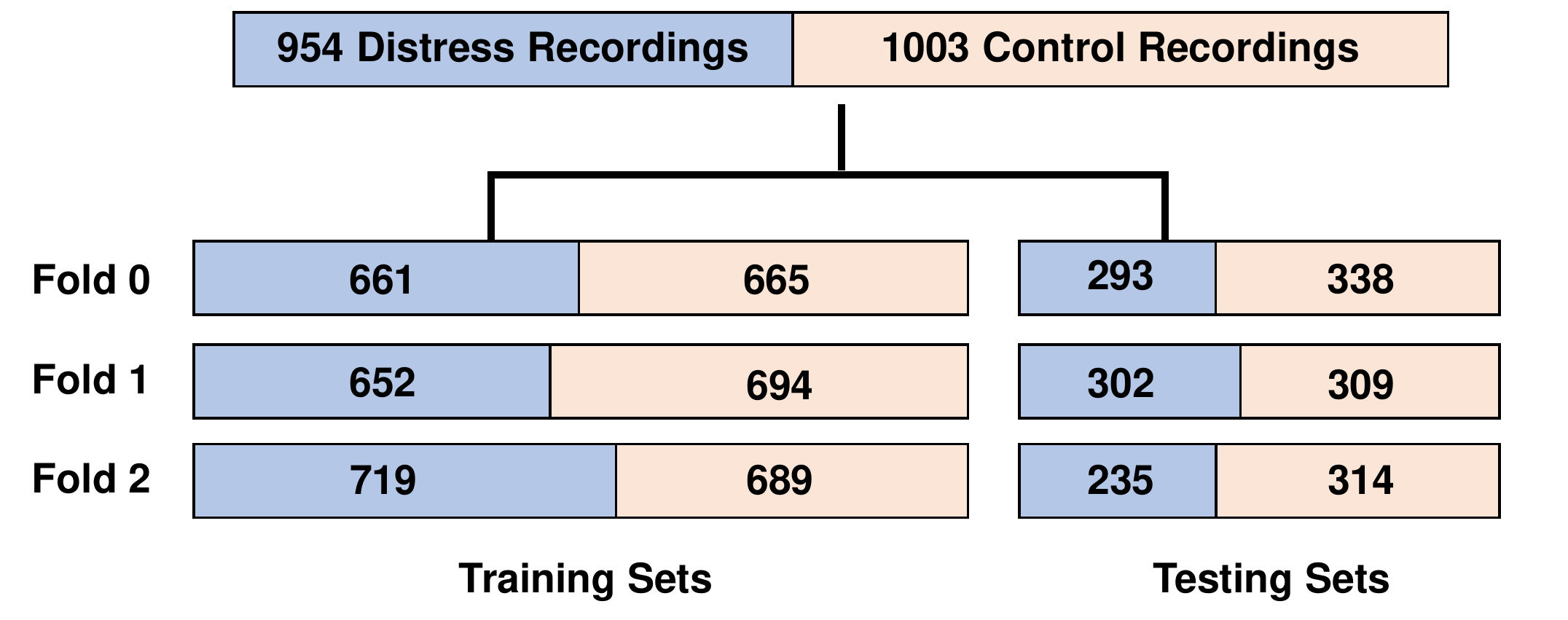}
  \caption{Summary of the dataset splits for the 3-fold cross validation experiments.}
  \label{fig:2}
\end{figure}

The entire 1957 segmented speech signals are distributed into 3 folds for cross-validation (Fig \ref{fig:2}). The train-test split for each fold is reserved to be approximately 70-30. We ensure that the same patient data is not used for both training and test so that the algorithms do not learn to recognize the speakers. 

\section{Proposed method}
\label{sec:method}

\subsection{Pre-processing}
The telephone calls typically include background noise and a static hum due to the recording instrument. Therefore, before feature extraction, we performed a speech enhancement operation to reduce the background noise.  

\subsection{Voice activity detection (VAD)}
The step is a pre-requisite for the extraction of the prosodic feature set. 
We use the VAD approach presented in \cite{sohn1999statistical} as provided in the MATLAB voicebox toolkit \cite{Brookes2005VoiceboxIA}. 

\subsection{Feature extraction}
\subsubsection{Acoustic features}\label{acoustic feat}
We use the Interspeech 2010 Paralinguistic Challange feature set proposed in \cite{is2010}. 
The feature set contains 38 low-level descriptors (LLDs) and 21 functionals yielding a total of 1582 acoustic features \cite{intaudio} as summarized in Table~\ref{tab:4}. 
The LLDs are extracted at 100 frames per second with a diverse set of short-time windows 
and smoothed by simple moving average low-pass filtering with a window length of 3 frames. Afterward, the first-order regression coefficients are calculated followed by the 21 functionals for every instance in the dataset \cite{intaudio}. These features are extracted using the openSMILE toolkit \cite{opensmile}.

\begin{table}[t]
  \caption{Acoustic Feature Set: MFCC: Mel-frequency cepstral coefficients, LSP: Line Spectral Pair, F\textsubscript{0}: Fundamental frequency, SHS: Sub- harmonic Summation}
  \label{tab:4}
  \centering
  \begin{tabular}{ll}
    \hline
    \multicolumn{1}{c}{\textbf{LLDs}} &  \multicolumn{1}{c}{\textbf{Functionals}} \\
    \hline
    Loudness                                &Rel. Pos. of max.,min.~~~                   \\
    MFCC [0-14]                             &Arith. Mean, Std. Deviation~~~              \\
    LogMelFreqBand [0-7]                    &Skewness, Kurtosis, Quart1,2,3~~~             \\
    LSP Frequency [0-7]                     &Inter-Quart. Range 2-1,3-2,3-1~~~            \\
    F\textsubscript{0} (SHS Based)          &Percentile 1\%,99\%~~~                      \\
    F\textsubscript{0} envelope             &Percentile Range 1\%-99\% ~~~               \\
    Probability of Voicing                  &\%Signal above min. +.9 range~~~            \\
    Jitter                                  &\%Signal above min. +.75 range~~~           \\
    Jitter of jitter                        &Linear Regression Coefficients~~~           \\
    Shimmer                                 &Linear Reg. Error Quad., Abs.~~~            \\
    \hline
  \end{tabular}
\end{table}

\subsubsection{Prosodic features}
We only utilize speaking rhythm related features in this work. 
The following four (4) LLDs are computed: (i) voice rate, (ii) voice duration, (iii) pause duration, and 
(iv) voice \& pause ratio. 
A window size of 1 sec is used and 21 functionals (as in Sec. \ref{acoustic feat}) are calculated to obtain an 84 dimensional feature set. 
 
\subsection{Feature normalization}
The extracted features have a high dynamic range and thus a normalization step is necessary for effective classification \cite{normimp}. In this study, we use standard mean and variance normalization for each feature. Normalization of both training and test features is done for each fold using the mean and variance calculated from the training data of the corresponding fold. 

\subsection{Machine learning classifier}
The SVM classifer has been used for training the dataset with binary decision using the LIBSVM \cite{libsvm} tool. 
The two classes include (i) distressed, and (ii) normal. The ``Distress" class contains data of patients who reported having mild or severe respiratory distress. The ``Normal" class contains the speech segments of healthy control subjects. 
A linear kernel is used for the SVM model with a fixed seed for every fold to ensure the results are reproducible. 

\section{Experimental Evaluation}
\label{sec:results}
The proposed method's performance is evaluated with respect to the accuracy, sensitivity, specificity, F1-score, and AUC (Area Under the Curve of ROC) for each fold. The mean and standard deviation of each performance metric is calculated from the folds. The acoustic and prosodic feature sets are first evaluated separately. In the final stage, the feature sets are fused by concatenation, and correlation-based feature reduction is performed to reduce the feature dimension to 251.  
The results are summarized in Table \ref{tab:6}. 

\begin{table}[t]
    \caption{Experimental Results over the 3-Fold Cross-Validation Data. Performance metrics include Accuracy (Acc.), Sensitivity (Sen.), Specificity (Spec.) F1-score (F1 Sc.) and Area Under the ROC Curve (AUC)}
    \label{tab:6}
    \small
    \centering
    \begin{tabular*}{0.48\textwidth}{ccccccc}
    \hline
    \begin{tabular*}{0.045\textwidth}[c]{@{}c@{}}\textbf{Feature}\\\textbf{Set}\end{tabular*} & \begin{tabular*}{0.045\textwidth}[c]{@{}c@{}}\textbf{Feat.}\\\textbf{size}\end{tabular*} & \begin{tabular*}{0.045\textwidth}[c]{@{}c@{}}\textbf{Acc.}\\\textbf{(\%)}\end{tabular*} & 
    \begin{tabular*}{0.045\textwidth}[c]{@{}c@{}}\textbf{Sen.}\\\textbf{(\%)}\end{tabular*} & 
    \begin{tabular*}{0.045\textwidth}[c]{@{}c@{}}\textbf{Spec.}\\\textbf{(\%)}\end{tabular*} & 
    \begin{tabular*}{0.045\textwidth}[c]{@{}c@{}}\textbf{F1 Sc.}\\\textbf{(\%)}\end{tabular*} &
    \begin{tabular*}{0.045\textwidth}[c]{@{}c@{}}\textbf{AUC}\\\textbf{(\%)}\end{tabular*}  \\ 
    \hline
    \begin{tabular*}{0.045\textwidth}[c]{@{}c@{}}Acoustic\end{tabular*} &  
    1957 &
    \begin{tabular*}{0.045\textwidth}[c]{@{}c@{}}86.4\\$\pm$2.05\end{tabular*} &        
    \begin{tabular*}{0.045\textwidth}[c]{@{}c@{}}85.9\\$\pm$3.5\end{tabular*}  &
    \begin{tabular*}{0.045\textwidth}[c]{@{}c@{}}86.9\\$\pm$0.5\end{tabular*}  &
    \begin{tabular*}{0.045\textwidth}[c]{@{}c@{}}86.0\\$\pm$0.57\end{tabular*} &   
    \begin{tabular*}{0.045\textwidth}[c]{@{}c@{}}92.0\\$\pm$2.08\end{tabular*} \\
    \hline
    \begin{tabular*}{0.045\textwidth}[c]{@{}c@{}}Prosodic\end{tabular*} &  
    84 &
    \begin{tabular*}{0.045\textwidth}[c]{@{}c@{}}56.5\\$\pm$2.51\end{tabular*} &        
    \begin{tabular*}{0.045\textwidth}[c]{@{}c@{}}57.8\\$\pm$8.88\end{tabular*}  &
    \begin{tabular*}{0.045\textwidth}[c]{@{}c@{}}54.7\\$\pm$6.19\end{tabular*}  &
    \begin{tabular*}{0.045\textwidth}[c]{@{}c@{}}58.4\\$\pm$3.42\end{tabular*} &   
    \begin{tabular*}{0.045\textwidth}[c]{@{}c@{}}59.0\\$\pm$2.08\end{tabular*} \\
    \hline
    \begin{tabular*}{0.045\textwidth}[c]{@{}c@{}}Fusion\end{tabular*} &
    251 &
    \begin{tabular*}{0.045\textwidth}[c]{@{}c@{}}85.8\\$\pm$0.95\end{tabular*} &        
    \begin{tabular*}{0.045\textwidth}[c]{@{}c@{}}85.6\\$\pm$4.15\end{tabular*}  &
    \begin{tabular*}{0.045\textwidth}[c]{@{}c@{}}86.3\\$\pm$3.96\end{tabular*}  &
    \begin{tabular*}{0.045\textwidth}[c]{@{}c@{}}86.6\\$\pm$0.53\end{tabular*} &   
    \begin{tabular*}{0.045\textwidth}[c]{@{}c@{}}92.0\\$\pm$2.65\end{tabular*} \\
    \hline
    \end{tabular*}
\end{table}

From the result, we observe that for acoustic features, the classifier shows a mean accuracy of 86.4 $(\pm 2.1)\%$ with the best sensitivity, specificity, F1 score, and AUC. The acoustic features have consistent performance across the 3 folds, including identical AUC values indicating the subject/patient invariance of the classifier. 
Compared to the acoustic features, the prosodic features performed were sub-par. This feature set resulted in an accuracy of only 56.5 $(\pm 2.5)\%$, which slightly better than a random classifier. The fusion of the features and feature selection did not provide any significant improvement in the overall performance. 
\begin{figure}[t]
  \centering
  \includegraphics[trim=15 0 30 30,clip,width=\linewidth]{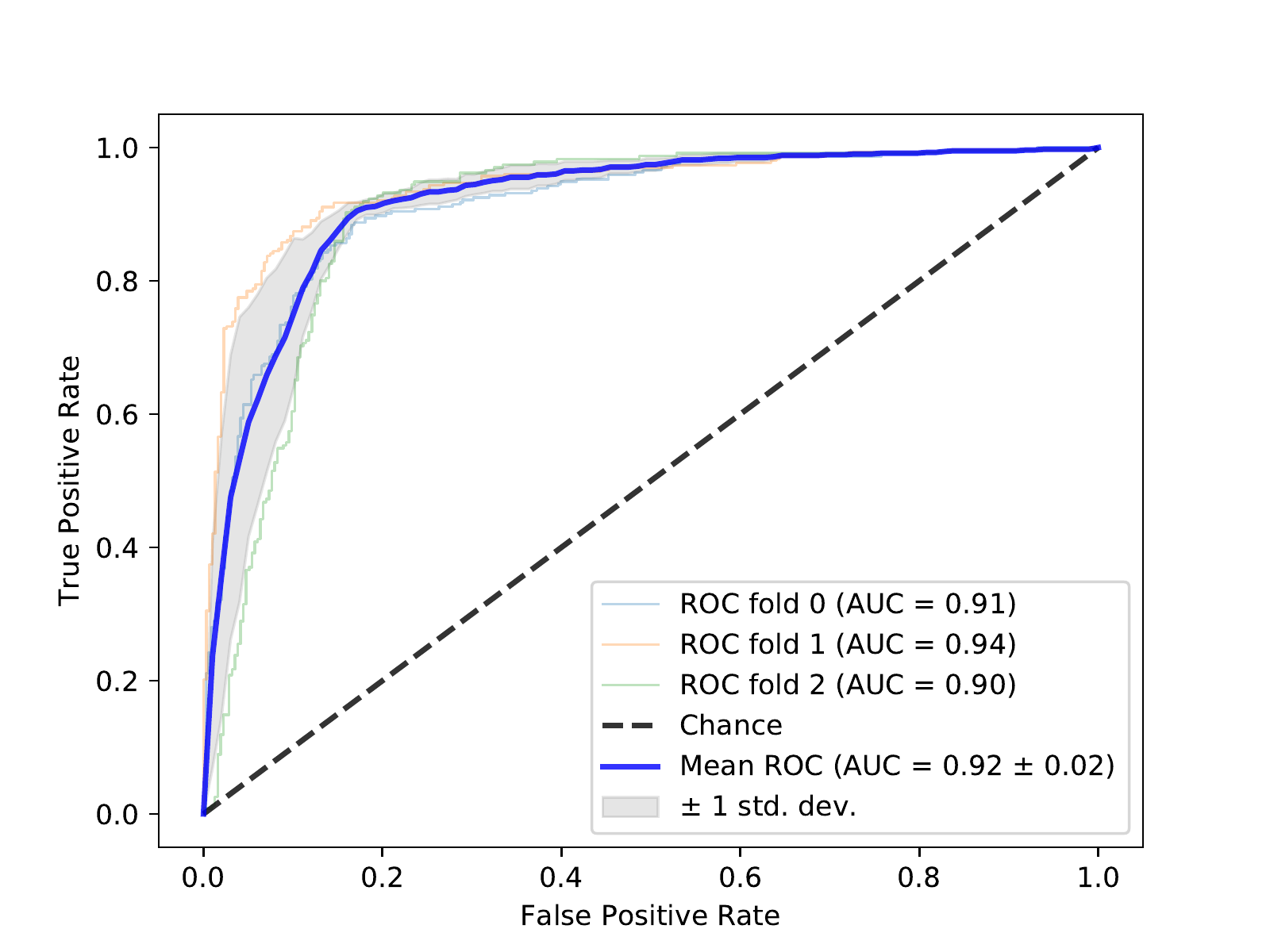}
  \caption{Receiver operating characteristics (ROC) curve for classification using acoustic features alone over the 3 folds.}
  \label{fig:5}
\end{figure}

\subsection{Analysis of feature ranking}
To correlate individual features' performance to the physiological aspects of speech production, we combined the acoustic and prosodic feature sets and performed a correlation-based feature ranking. The analysis returned 251 top-ranked features, including 242 acoustic features and 9 prosodic features. To simplify the study, we look at the LLD features included among the top 251 features and provide the rankings of the top 10 LLDs in Table \ref{tab:rank}. We observe that features that can be physiologically connected to respiratory distress are included, such as loudness, voice rate, voice duration, pause duration. Loudness is the most prominent feature in detecting respiratory distress. Spectral and cepstral features are also found to be important in the ranking analysis, confirming our hypothesis that voice quality can be altered due to respiratory distress (e.g., increased hoarseness).  
\begin{table}[ht]
  \caption{LLD Ranking}
  \label{tab:rank}
  \centering
 
  \begin{tabular}{lc}
    \hline
    \multicolumn{1}{c}{\textbf{LLD Name}} &  \multicolumn{1}{c}{\textbf{Rank}} \\
    \hline
    Loudness                    &1             \\
    MFCC                        &2            \\
    logMel Freq. Band              &3\\
    Probability of Voicing           &4\\
    Pause Duration              &5\\
    Voice Rate                  &6\\
    lsp Frequency                &7\\
    Jitter                      &8\\
    Voice Duration              &9\\
    Shimmer                     &10\\
    \hline
  \end{tabular}
\end{table}

\section{Conclusion}
\label{sec:conclusion}
This study presented findings from our preliminary analysis on respiratory distress detection from telephone speech collected from real telemedicine phonecalls from actual patients reporting adverse health conditions. We have utilized a set of acoustic and prosodic features for binary classification between a patient reporting respiratory distress and a healthy control subject.
Experimental results using the proposed feature set and an SVM classifier showed promising results achieving above 85\% performance in all of the performance metrics, namely, accuracy, sensitivity, specificity, and F-1 score, on a 3-folds cross-validation experiment. The top-ranked features obtained by correlation analysis were found to be physiologically meaningful.
The proposed method can significantly impact the automatic early detection of respiratory diseases through telemedicine phonecalls in low-resource settings.
A further large-scale clinical study is required to confirm the diagnostic usability of the proposed approach.

\vfill\pagebreak

\small
\bibliographystyle{resdistress_prediction}

\bibliography{resdistress_prediction.bib}

\end{document}